\documentclass{epl}

\title{Vortex correlations in a fully frustrated two-dimensional superconducting 
 network}
\shorttitle{Vortex correlations}
\author{E. Serret, P. Butaud and B. Pannetier}
\institute{CNRS-CRTBT, associ\'e \`a l'Universit\'e Joseph Fourier\\
25 Av.  des Martyrs, 38042 Grenoble Cedex 9, France
}

\begin{document}

\maketitle

\begin{abstract}
We have investigated the vortex state in a superconducting dice
network using the Bitter decoration technique at several magnetic
frustrations $f=\phi / \phi_{0}$=1/2 and 1/3.  In contrast to
other regular network geometries where the existence of a commensurate
state was previouly demonstrated, no ordered state was observed in 
the dice network at $f=1/2$ and the observed vortex-vortex correlation 
length is close to one lattice cell.
\end{abstract}

\section{Introduction}
In the past decades, the vortex state of superconducting networks has
been investigated by several groups using different imaging techniques
(scanning Hall microscopy \cite{Hess,Chang}, scanning SQUID
microscopy\cite{VanHarlingen} or Bitter decoration
\cite{Runge,Bezryadin,Eichenberger}).  The vortex configuration was
studied in square or triangular lattices as a function of the magnetic
field.  In superconducting arrays the relevant variable is the
magnetic frustration, $f$, which represents the vortex filling factor. 
$f$ is defined as $f=\phi / \phi_{0}$ with $\phi_{0}=h/2e$, the flux
quantum, and $\phi$, the magnetic flux per elementary plaquette. 
The vortex pattern reflects the spatial phase configuration of the
superconducting order parameter resulting from the competition between
the magnetic field and the underlying lattice \cite{Pannetier84}.  For
rational frustration $f=p/q$, with $p$ and $q$ integer numbers, a
commensurate ground state is generally expected as for example the
checkerboard state which was imaged in square lattices at $f=1/2$
\cite{Hess,Runge}.

The purpose of this Letter is to present the detailed results of a
magnetic decoration experiment in a fully frustrated dice network. 
The unusual properties of the dice lattice were recently highlighted
by the discovery of a peculiar destructive interference phenomenon
occurring at $f=1/2$ for the electronic wave function in the
tight-binding model\cite{Vidal1,Vidal2}.  This phenomenon was shown
later to manifest itself as a broadening of the superconducting
transition and a suppression of the critical current in the
corresponding superconducting wire network\cite{Abilio,Naud}.  This
observation was indicative of the absence of commensurate vortex state
in the dice network, in contrast to the square network which was shown
to exhibit a sharp critical current peak at $f=1/2$\cite{Buisson}.  We
report here new careful decoration experiments on a series of high
quality extended niobium dice networks under different magnetic
fields.  The analysis of the vortex correlation functions shows
unambiguously that the vortex state is fully disordered at $f=1/2$ and
confirms the results suggested by our preliminary experiments on small
networks \cite{ms2000}.

\begin{figure} 
\onefigure[width=4cm]{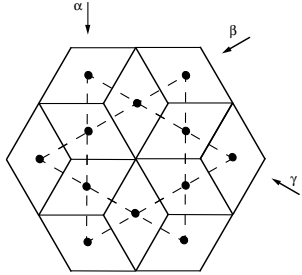}

\caption{The dice lattice is shown here as full lines.  Black
points at the center of the cells, illustrate the location
of vortices at the nodes of the Kagom\'e lattice symbolized by dashed
lines. Arrows show the 3 directions along which the correlation functions 
(Eq.\ref{e.1}) have been studied.}

\label{Kagome}
\end{figure}

The issue of the vortex configuration is of additional interest in the
dice lattice as the vortices are located on the (dual) Kagom\'e
lattice (Fig.  \ref{Kagome}).  Since the vortex variable is binary,
for $f$ smaller than 1 (in each cell there is either one vortex ($+1$)
or no vortex ($-1$)), we can view this problem as Ising spins on the
nodes of a Kagom\'e lattice.  This general problem has been studied
theoretically\cite{Mekata,Moessner} with first and second neighbour
interaction.  Our vortex experiment is the first observation of binary
variables on a the Kagom\'e lattice.

\section{Samples and experimental technique}
\begin{figure} 
\twoimages[height=5cm]{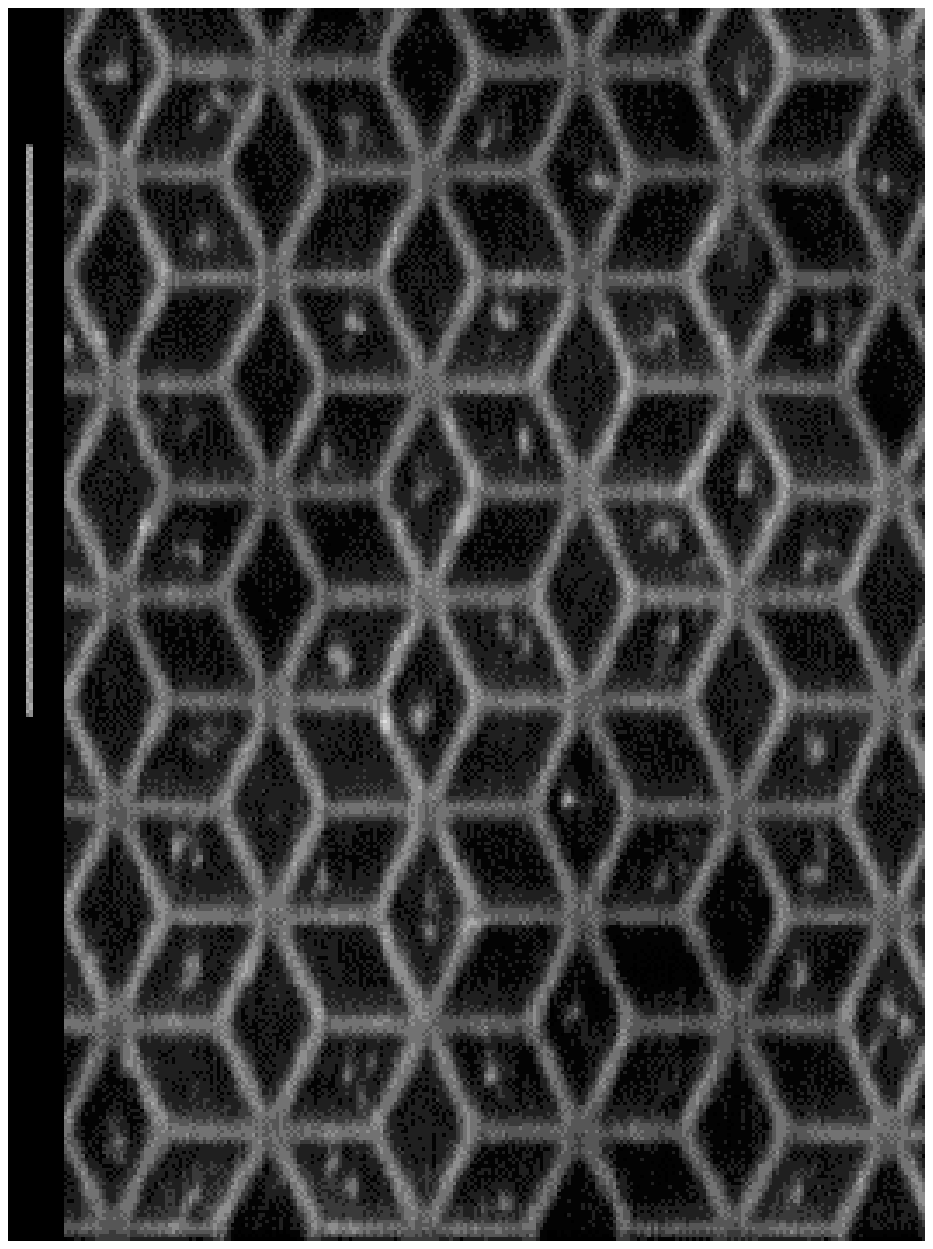}{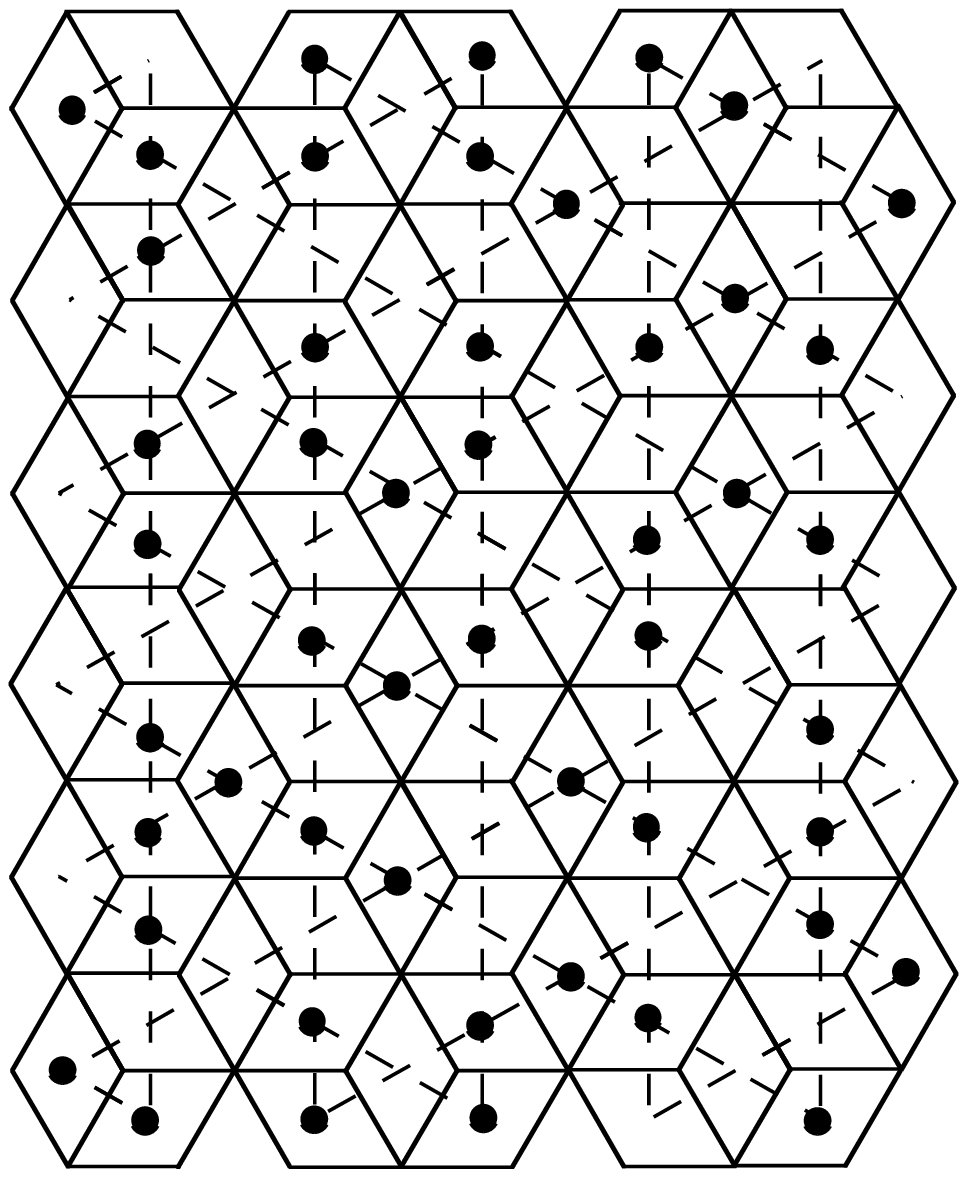}
\caption{Left : Partial view of the imaged network after
decoration at $f=1/2$.  The length of each wire is 1 $\mu$m. The vertical bar on the 
left corresponds to 5 $\mu$m. The white points
visible near the center of the cells are the images of the Ni clusters
that decorate vortices.  Right : the transcription of the vortex
configuration.  The black dots symbolize the observed vortices.  The
dice lattice is represented with full lines and the Kagom\'e (dual 
lattice) with dashed lines.}

\label{pattern}
\end{figure}

 The imaging was performed by the Bitter magnetic decoration technique
 \cite{Trauble}.  Since the magnetic contrast of vortices in
 superconducting wire networks\cite{Runge} is very weak, we used the
 so-called flux compression technique\cite{Bezryadin}.  The networks
 (Fig.  \ref{pattern}) were made of niobium wires ($1 \mu$m long,
 $100$ nm wide) patterned from a $200$ nm thick Nb layer epitaxially
 grown on a sapphire substrate at 550$^{\circ}$C. The patterning was
 achieved by reactive ion etching in $SF_{6}$ through an aluminum mask
 prepared by \textit{e-beam} lithography\cite{Leica}.  Only the first $130$ nm
 Nb was removed, leaving unpatterned a $70$ nm thick niobium layer as
 a flux compression bottom layer.  The total array size was $800 \mu$m
 long and $600 \mu$m wide.  The networks contained about $550 000$
 cells, with a wire width homogeneity better than $10 \%$.  The
 elementary cell area in all networks is $0.866\mu$m$^{2}$ and the
 matching field ($f=1$) was obtained for $B=2.39$ mT. The decoration
 cell was placed inside a double $\mu$-metal shield.  Before each run
 the magnetic field was calibrated using the magnetoresistance curve
 of a large SNS Josephson junction array.  The flux accuracy is a few
 $10^{-3}\phi_{0}$ per cell.  We always found the counted vortex
 density perfectly consistent with our field calibration.  We first
 apply the magnetic field in the normal state at $10$ K. Then the
 sample is slowly cooled down across the niobium transition
 temperature.  Vortices nucleate at the niobium superconducting
 transition temperature, $T_{c}=9.0$ K for our samples, and, because
 of the strong network pinning, their configuration freezes out as the
 temperature decreases a few mK below $T_{c}$\cite{remark}.  Flux
 compression takes place at the superconducting transition
 ($T_{c}=8.93$ K) of the bottom layer when the network vortex loops
 convert into Abrikosov vortices.  Once the temperature is stabilized
 at $4.2$ K, Ni particles are flash-evaporated on the sample under a
 nominal residual helium pressure of $0.6$ mbar.  The vortex positions
 are then registrated at room temperature using a scanning electron
 microscope (cf Fig.  \ref{pattern}).  This "one-shot" technique
 allows us to image a wide region and collect statistical information
 on the position of vortices.

\section{Experimental results}
\begin{figure} 
\onefigure[width=12cm]{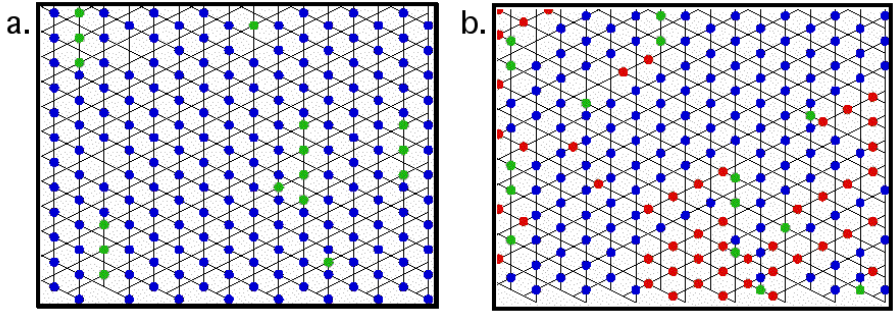}
\caption{A part of the configurations observed in two different 
decoration runs : $f=1/3 + 0.005$ (a) and $f=1/3+0.016$ (b). The circles on 
the nodes of the Kagom\'e lattice represent the positions of the observed 
vortices. The blue, green and red colors indicate the three 
different positions in the Bravais cell. Single colored domain 
correspond to one of the three equivalent commensurate states which are 
degenerate because of the 3-fold symmetry.}
\label{configtiers}
\end{figure}

\begin{figure} 
\onefigure[width=13cm]{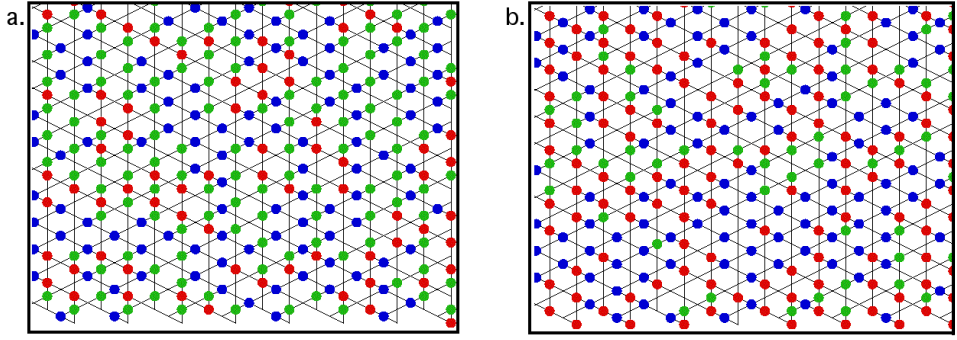}
\caption{A part of the
observed configurations observed in two different samples for 
$f=1/2$. The effective filling is (a) $f=1/2-0,0014$ and (b) 
$f=1/2+0,0205$.The colored circles represent the position of the vortices on nodes of
the Kagom\'e lattice. Blue, green and red colors correspond to the 
three positions on each dice.}
\label{configdemi}
\end{figure}

In this section we compare the vortex structures at $f=1/3$ and at
$f=1/2$.  The datas are extracted from the SEM-micrographs and the
vortex configuration is then regenerated on a computer.  The pictures
on Fig.  \ref{configtiers} and \ref{configdemi} are parts of the
complete images.  Since the Bravais cell is composed of three
different plaquettes, we represent the vortices with three different
colors (blue, green, and red) depending on where they are located.  To
obtain a quantitative information on the degree of disorder, we have
calculated linear correlation functions between vortex variables in
the three equivalent directions of the lattice (see Fig. 
\ref{Kagome}).  By making distinction between thoses three directions,
we keep information on the domain shapes, which would be lost on
averaging over the three contributions.  The vortex variables $V_{i}$
are equal to $+1$ if a vortex is present in the $i$ cell and $-1$ if
not.  The three correlation functions, $C_{\alpha}$, $C_{\beta}$ and
$C_{\gamma}$ are defined as :

     \begin{equation}
     \label{e.1}
     C_{\alpha,\beta,\gamma}(r) = \langle V_{i}.V_{i+r} \rangle _{\alpha,\beta,\gamma}
     \end{equation}

\noindent where the $i+r$ cell is the $r$-th neighbour of the $i$ cell
in the $\alpha$, respectively $\beta$ or $\gamma$, direction (see Fig. 
\ref{Kagome}) and $\langle\ldots\rangle_{\alpha,\beta,\gamma}$ is the mean product
$V_{i}.V_{i+r}$ for each $i$ having at least one $r$-th neighbour in the
$\alpha$ (respectively $\beta$ or $\gamma$) direction.

\begin{figure} 
\twoimages[width=6cm]{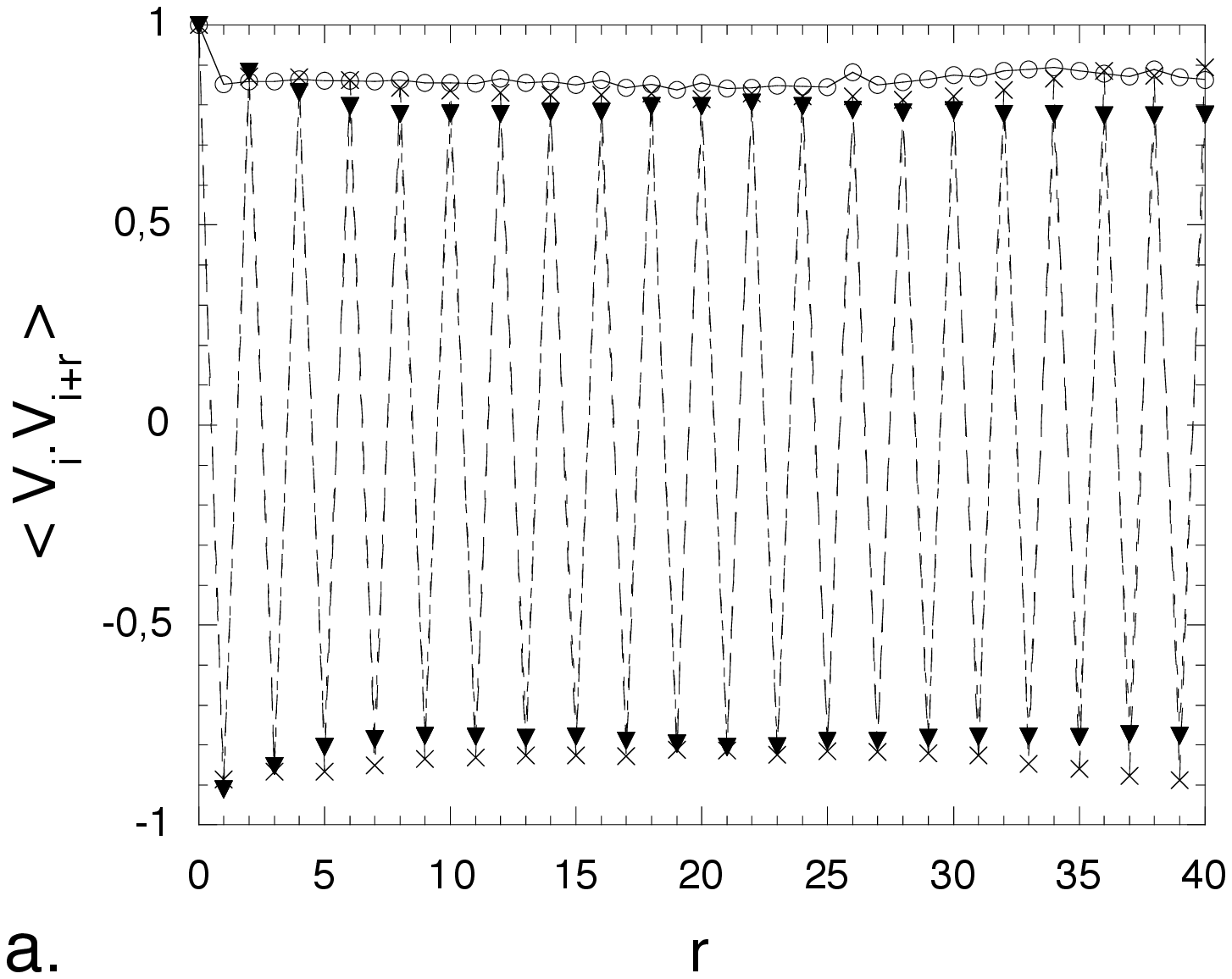}{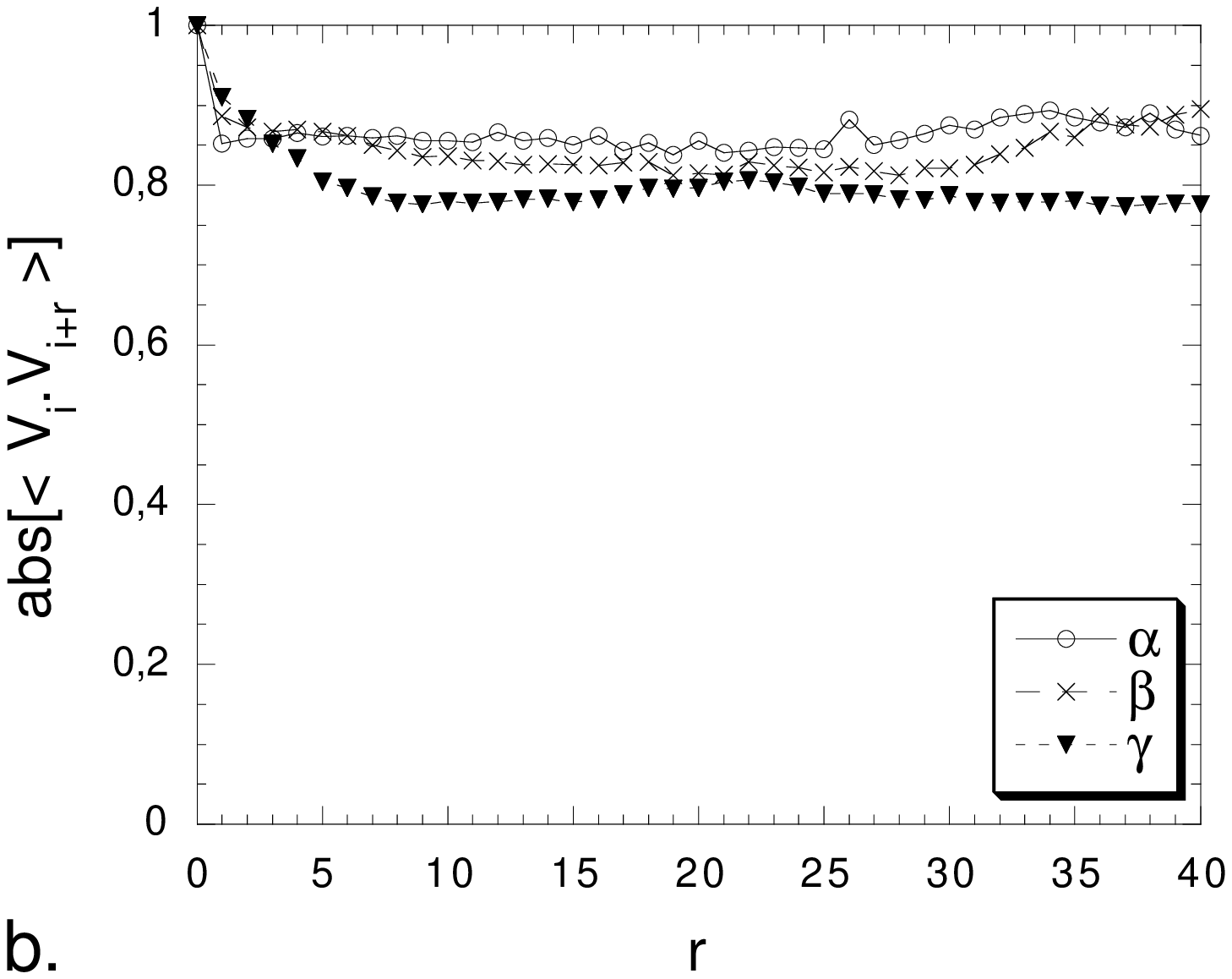}
\caption{a: correlation functions on the vortex 
position at $f=1/3 + 0.005$. The lines are guide for the eyes since 
the function is only defined for entire r. The correlation function exhibits 
a kind of antiferromagnetic order in the $\beta$, $\gamma$ directions 
(1,-1,1,-1,1\ldots) and a kind of ferromagnetic order in $\alpha$
direction  (1,1,1,1,1\ldots or -1,-1,-1,-1\ldots). The absolute value 
 of correlation (figure b) is still about 0.85 at $r=40$. }
\label{corretiers}
\end{figure}

We first consider filling factor $1/3$ where a stable commensurate
state is expected.  Indeed one can build a simple ordered configuration with 1
vortex every $3$ cells, i.e. one vortex per Bravais cell.  This
ordered phase is three times degenerated due to the lattice symmetry. 
On Fig.  \ref{configtiers}, we show two of those phases observed on
two different runs performed at magnetic field $B=0.797$ mT. We have
analyzed each time a zone of approximately $3000$ cells at the
network's center.  The actual frustration, as determined from the
observed number of vortices, is for the first run: $f=1/3 + 0.005 $
(Fig.  \ref{configtiers} left).  We observe a huge single domain
extending over the whole picture.  For the second run shown in Fig.
\ref{configtiers} right, the actual frustration is $f=1/3 + 0.016 $. 
The observed effective domain size is approximately 10 cells.  Excess
vortices seem to gather near the domains walls, as can also be noticed
in the former case (added vortices sit preferentially on the defect
lines).  This behavior is similar to the one observed in square arrays
\cite{Hess}.  The correlation functions for $f=1/3 + 0.005$ are shown
on Fig.  \ref{corretiers}.  As can be seen, the correlation falls from
$1$ at $r=0$ to approximately $0.85$ at $r=40$ in the three directions
and this level of correlation is conserved beyond the 40-th neighbour. 
The extended ordered vortex state at $f=1/3$, is similar to that
observed in the square or triangular lattices with rationnal
frustrations \cite{Hess,Runge}.  This structure remains ordered even
within a few percent of excess vortices but the domain size decreases. 
This indicates the robustness of the $1/3$-ordered state to extra
vortices.

\begin{figure} 
\twoimages[width=6cm]{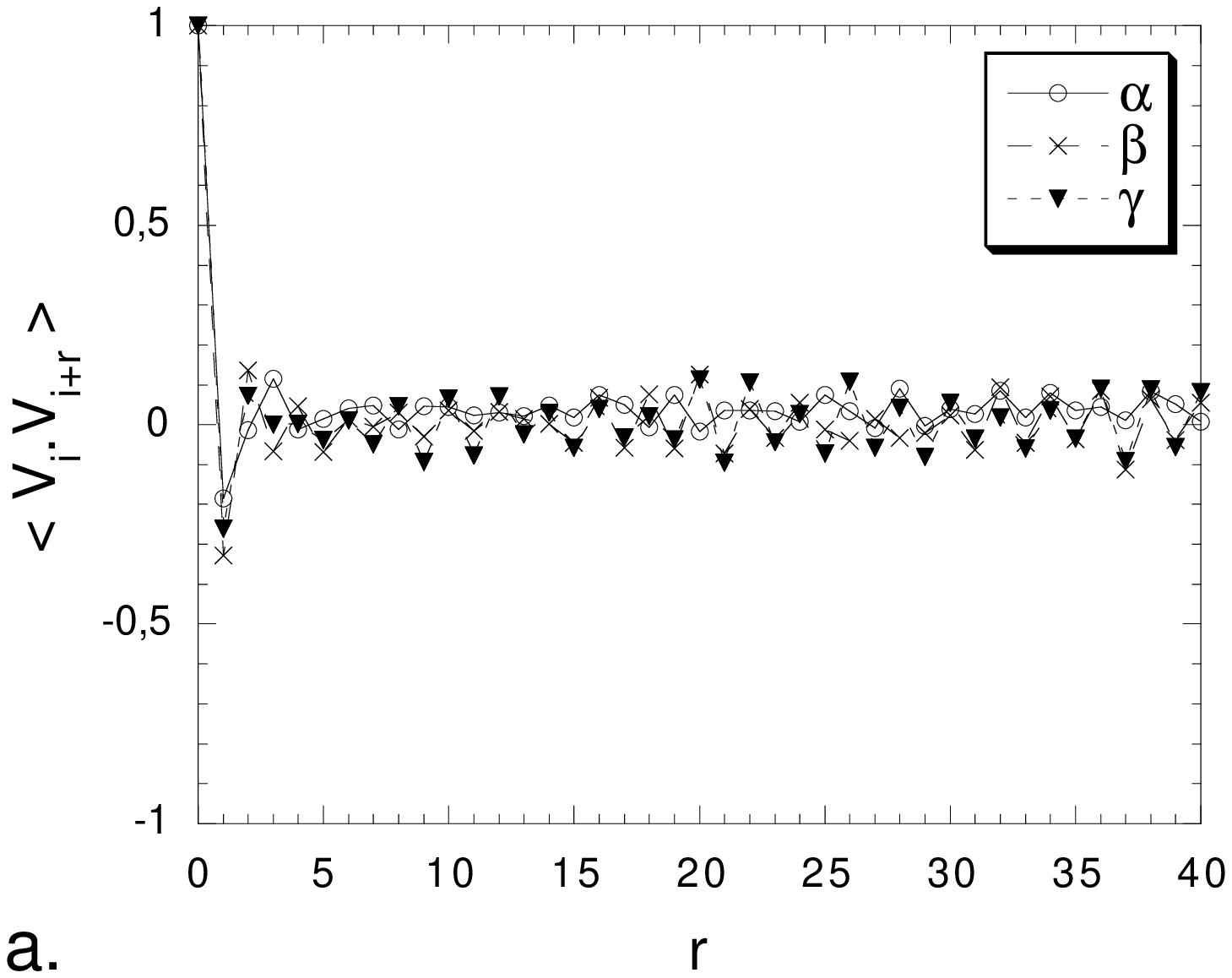}{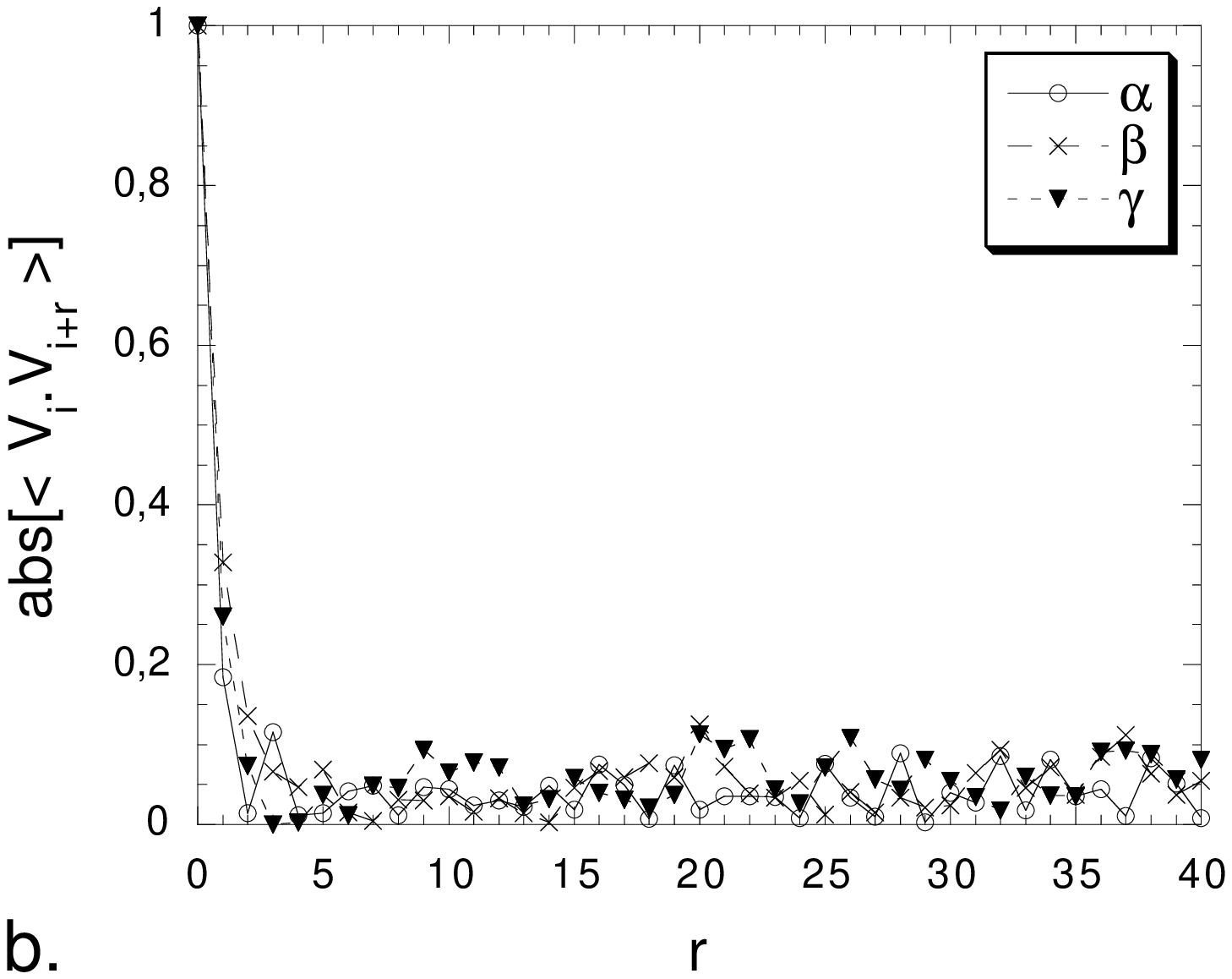}
\caption{a: Correlation functions on the vortex position at 
$f=1/2-0,0014$. b: Absolute value of the correlation functions. We clearly 
see a drop of the correlation function in the three directions at approximately $r=3$ with 
$f=1/2$.}
\label{corrdemi}
\end{figure}

In contrast, at half filling the vortex lattice is incommensurate with
the dice lattice.  One must consider two Bravais cells.  We performed
three different experiments at $f=1/2$.  The number of investigated
cells was approximately $4000$ for each experiment.  The measured
frustration was $\delta f= -0.0014, +0.0015, +0.0205$ around the ideal
half filling.  A part of two regenerate images is displayed on Fig. 
\ref{configdemi}.  Despite the fact that $f=1/2$ is a rational filling
factor, we did not observe simple ordered phases in any of the three
experiments.  Rather we observe strongly disordered states reminiscent
of the incommensurate states occuring at irrational frustrations \cite
{Halsey}.  The correlation functions (see Fig.  \ref {corrdemi})
rapidly drops to approximately $0.05$ beyond $r=2$.  This residual
correlation is a real effect, larger than the experimental 
uncertainties. It is due to differences in the probability of populating the
three different sites in the Bravais cell. It corresponds for this 
decoration experiment to a 
population ratio of $0,62$, $0,32$ and $0,56$ respectively for the 
different kind of sites. This difference could be
due to some slight inhomogeneity in the network temperature during cooling or
in the wire geometry\cite{defects}.  Presently the origin of this small effect is
not understood and remains an open problem.  Fitting the correlation function with an exponential
function $\propto e^{-r/\xi}$, we get a correlation length $\xi
\approx 1.5$ for $f=1/2$.

We also imaged a vortex lattice at $f=1/6 + 0.0082$.  We saw no
ordered domains, but we could not make definitive conclusion on the
$1/6$-state with this experiment.  This small deviation from the exact
$f=1/6$ indeed corresponds to $4.7 \%$ of the vortex population and
might change the configuration more efficiently than for a higher
filling factor such as $f=1/3$ or $f=1/2$.  In the litterature, no
observation of ordered vortex state was ever reported in superconducting
networks for such small filling factors.

\section{Discussion}
Our observation of a very short correlation length at $ f = 1/2 $, is
indicative of a strongly disordered state.  This observation is
consistent with previous transport measurements in superconducting Al
networks with the same lattice\cite{Abilio}.  A sharp peak was
observed at $f=1/3$, consistent with the strongly pinning commensurate
state shown in Fig.  \ref{configtiers}.  On the other hand the
suppression of critical current at $f=1/2$ was indicative of a
weakening of the phase rigidity due to the localization of the
superconducting wave function in the Aharonov Bohm cages\cite
{Vidal1}.  The absence of ordering is reminiscent of the problem of
classical Ising spins (zero or one vortex per cell) on the nodes of
the Kagom\'e lattice\cite{Huse}.  However, since the condition of nearest
neighbour interaction which is central to the prediction of disordered
state at all temperatures in the Ising model, is presumably not
fulfilled for array vortices one cannot draw conclusions from this
analogy.  Very recent theoretical studies of vortices in dice Josephson junctions
arrays have more relevance as they directly address this issue of 
vortex ordering.

S.E. Korshunov \cite{Korshunov} investigated the ground state of
superconducting dice arrays with either cosine or quadratic dependence
of the coupling energy upon the phase difference.  He found a ground
state made of an ordered configuration of vortex triads (clusters of
three adjacent vortices).  Because of the proliferation of zero-energy
domain walls, the ground state entropy is large but it is not
extensive, in contrast to the case of Ising spins on a Kagome
lattice\cite{Huse} where an extensive entropy was found.  Therefore,
in this model, the true ground state configuration for $f=1/2$ should
be ordered.  In contrast, the observed vortex correlation functions in
our decoration experiments do not reveal any signature of order. 
Also, the histogram of vortex cluster sizes shows no preferred
occurrence of triads.  We do not believe that the geometrical
irregularities present in experimental sample\cite{defects} are
responsible for the observed disordered state.  Instead, we believe
that it results from the combination of large thermal fluctuations and
large entropy of low energy configurational states at the vortex
freezing temperature\cite{remark}.  Indeed, a large number of low
energy vortex configurations are likely to exist in the dice lattice
\cite{Feigel'man,Fazio}.  However, their difference in energy is much less
than the energy barrier for vortex motion which, in the wire
network\cite{Giroud} is mostly determined by the single wire length,
independently of the network topology.  On cooling down the vortex
lattice may therefore be freezed out in a configuration different from
the true ground state.  A possible origin of the lack of ordering of
the vortex lattice was recently pointed out by V. Cataudella and R.
Fazio\cite{Fazio}.  Their Monte Carlo simulations on the fully
frustrated Josephson junction dice array suggest the existence of a
low temperature phase transition below which a glassy dynamics
prevents the system to reach the true ground state conjectured in
Ref\cite{Korshunov}.  This scenario is consistent with our observation
but cannot be tested in our strong vortex pinning wire network array.

\section{Conclusion}

We performed magnetic imaging of the vortex configuration in Nb dice
networks under several magnetic fields.  We focused our study on
magnetic frustrations $f=1/3$ and $f=1/2$.  Numerical calculations of
the correlations on the observed vortex positions lead to a
quantitative characterization of the degree of disorder.  While a
robust commensurate state is found at $f=1/3$, we observe a strongly
disordered vortex configuration at $f=1/2$ characterized by a vortex
correlation length close to one cell size.  It is worth noticing that
this feature, particular to the dice lattice, appears here in a system
with no geometrical disorder.  The discrepancy with the Korshunov
prediction that the ground state is ordered suggests that the true
ground state is not accessible in our experiment.  Further
investigation of Josephson junction arrays on a dice lattice  by 
transport and high sensitive magnetic microscopy\cite{Veauvy} at much
lower temperature are needed to reveal the true nature of the ground
state.

\acknowledgments We acknowledge fruitful discussions with C.C. Abilio,
B. Canals, B. Dou\c{c}ot, S.E. Korshunov, M. Feigel'man, L. Ioffe, P. Martinoli,
R. Mosseri, R. Fazio and J. Vidal.  Discussions within the ESF "Vortex Matter
Program" and the TMR network No. FMRX-CT97-0143 are gratefully acknowledged. 
The \textit{e-beam} lithography and the SEM observations were carried
out with CEA-LETI-PLATO organization teams and tools.


\begin{thebibliography}{0}
\bibitem{remark} Vortex pinning is strong in wire arrays\cite{Giroud}
due to the large energy barrier for vortex crossing a superconduction
wire.  This barrier is suppressed in the vicinity of $T_{c}$ when the
superconducting coherence length which defines the vortex core size
becomes of order of the wire length.  The condition $\xi (T) \approx
a$ gives an indication of the onset of pinning by the wire array and
determines the freezing temperature of the vortex lattice.  It is
reached at $\approx 4$ mK below $T_{c}$.
\end{thebibliography}
\end{document}